\documentstyle[aps,epsf]{revtex}
\input epsf.sty
\def\prb{Phys. Rev. B}
\def\prl{Phys. Rev. Lett.}

\def\be{\begin{equation}}
\def\ee{\end{equation}}
\def\ba{\begin{eqnarray}}
\def\ea{\end{eqnarray}}

\def\YBCO{YBa$_2$Cu$_3$O$_{7-\delta}$}

\def\C60{A$_x$C$_{60}$}

\def\hty{high temperature superconductivity}
\def\hts{high temperature superconductors}
\def\ua{\uparrow}
\def\da{\downarrow}
\begin{document}
\draft
\flushbottom
\twocolumn[
\hsize\textwidth\columnwidth\hsize\csname @twocolumnfalse\endcsname

\title{Microscopic Theory of High Temperature Superconductivity}
\author{V.~J.~Emery$^1$ and S.~A.~Kivelson$^2$}
\address{
$^1$Dept. of Physics
Brookhaven National Laboratory
Upton, NY  11973-5000}
%}
%\author{S.~A.~Kivelson}
\address{
$^2$Dept. of Physics
University of California at Los Angeles
Los Angeles, CA 90095}
\date{\today}
\maketitle
\tightenlines
\widetext
\advance\leftskip by 57pt
\advance\rightskip by 57pt

\begin{abstract}

It is argued that the BCS many-body theory, which is outstandingly successful 
for conventional superconductors, does not
apply to the {\hts} and that a realistic theory must take account of the
local electronic structure (stripes). The spin gap proximity effect is
a mechanism by which the charge carriers on the stripes and the spins
in the intervening regions acquire a spin gap at a relatively high
temperature, with only strong repulsive interactions. 
Superconducting phase order is achieved at a lower
temperature determined by the (relatively low) superfluid density 
of the doped insulator. This picture is consistent with the phenomenology of 
the {\hts}. It is shown that, in momentum space, the spin gap first arises 
in the neighborhood of the points $(0,\pm \pi)$ and $(\pm \pi, 0)$ and then
spreads along arcs of the Fermi surface. Some of the experimental consequences
of this picture are discussed.

%\vskip 1cm

\end{abstract}

]

%\narrowtext
\narrowtext
\tightenlines

\section{Introduction}

The high temperature superconductors \cite{BM} are quasi-two dimensional 
doped insulators, obtained by chemically 
introducing charge carriers into a highly-correlated antiferromagnetic 
insulating state. There is a large ``Fermi surface'' containing all of the 
holes in the relevant Cu(3d) and O(2p) orbitals \cite{arpes}, but $n/m^*$ 
vanishes as the dopant concentration tends to zero.\cite{optical,muon} 
(Here $m^*$ is the effective mass of a hole and $n$ is either the superfluid 
density or the density of mobile charges in the normal state.) 
Clearly, understanding the origin of {\hty} and the nature of the doped 
insulating state are intimately related.

The doped insulating state is well understood in one dimension: the added 
charges form extended objects, or solitons, which 
move through a background of spins that have distinct dynamics.\cite{review} 
(This is the origin of the concept of the separation of spin and charge.)
In two dimensions the doped-insulating state also is characterized by a 
one-dimensional array of extended objects, but they are 
slowly-fluctuating, metallic charge stripes that separate the spins into 
antiphase domains. These self-organized structures are driven by 
the tendency of the correlated antiferromagnet to expel the doped holes, and 
not by specific features of the environment of the CuO$_2$ planes.\cite{bian}
The evolution of these ideas and the extensive evidence for this local 
electronic structure of the CuO$_2$ planes is described in a companion paper 
at this conference.\cite{rome1} 

Rather general and phenomenological arguments indicate that the BCS many-body 
theory, which is so successful for conventional superconductors, must be 
revised for the {\hts}. (Section II.) Once this is accepted, it is clear that 
any new many-body theory must be based on the local electronic structure of 
the doped insulator, especially structure on the scale of the superconducting 
coherence length. In Sec. III it will be shown that, locally, the stripe 
structure may be regarded as a quasi one-dimensional electron gas in an 
active environment provided by the antiphase spin domains. For a quasi 
one-dimensional system there are two routes to superconductivity --- a 
low-T$_c$ route that is analogous to BCS theory and a potentially high-T$_c$ 
route in which a spin gap is formed at a relatively high temperature and is 
independent of the onset of phase coherence which takes place at a lower 
temperature that is governed by the superfluid density. \cite{LE,sak}.
In a quasi-one-dimensional electron gas (1DEG), both routes require some 
sort of attractive interaction.\cite{review} However, the active environment adds 
a new element to the picture by allowing the formation of a spin gap with purely 
repulsive interactions via the ``spin-gap proximity effect.''\cite{sgpe} The 
driving force is a lowering of the zero-point kinetic energy of the mobile 
holes, and it constitutes our mechanism of {\hty}. In this way, the stripe 
picture allows us to derive the phenomenology of the {\hts}.

The symmetry of the order parameter emerges once these ideas are re-expressed
in momentum space. (Section IV.) It will be shown that $d$-wave symmetry gives
the lowest energy if the range of the gap function in real space is one
lattice spacing. However, second and third neighbor components of the gap 
function favor $s$-wave symmetry and, in certain circumstances, they could 
either mix with the $d$-wave component (breaking time-reversal symmetry or 
lattice-rotational symmetry) or even become dominant.

\section{BCS many-body theory}

It has been argued that the quasiparticle concept does not apply to many 
synthetic metals, including the {\hts}.\cite{badmetals} This idea is
supported by angular resolved photoemission spectroscopy (ARPES) 
on the {\hts}, which shows no sign of a normal-state quasiparticle peak near 
the points $(0,\pm \pi)$ and $(\pm \pi,0)$ where {\hty} 
originates.\cite{arpes3} If there are no quasiparticles, there is no Fermi
surface in the usual sense of a discontinuity in the occupation number
$n_{{\vec k}}$ at zero temperature. This undermines the very foundation of 
the BCS mean-field theory, which is a Fermi surface instability that
relies on the existence quasiparticles. 

A major problem for any mechanism of {\hty} is how to achieve a high pairing 
scale in the presence of the repulsive Coulomb interaction, especially in
a doped Mott insulator in which there is poor screening. 
In the {\hts}, the coherence length is no
more than a few lattice spacings, so neither retardation, nor a
long-range attractive interaction is effective in overcoming the bare 
Coulomb repulsion. Nevertheless ARPES experiments \cite{darpes} show that 
the major component of the energy gap is
$\cos k_x - \cos k_y$. Since the Fourier transform of this quantity vanishes
unless the distance is one lattice spacing, it follows that
the gap (and hence, in BCS theory, the net pairing force) is a maximum for holes 
separated by one lattice spacing, where the bare Coulomb interaction is very 
large ($\sim$ 0.5 eV, allowing for atomic polarization). It is not easy to 
find a source of an attraction that is strong enough to overcome the Coulomb
force at short distances {\it and} 
achieve {\hty} by the usual Cooper pairing in a natural way. 

Thus, although the outstanding success of the BCS theory for conventional 
superconductors tempts us to use it for the {\hts}, it is clear that we 
should resist the temptation and seek an alternative many-body theory.
There is phenomenological support for this point of view. In the BCS mean-field 
theory, an estimate of T$_c$ is given by T$_c \sim \Delta_0 /2$, where $\Delta_0$ is 
the energy gap measured at zero temperature. This is a good approximation
for conventional superconductors because the classical phase ordering 
temperature $T_{\theta}$ is very high. A rough upper bound on  $T_{c}$ is 
obtained by considering the disordering effects of 
only the classical phase fluctuations as
$T_{c}\sim T_{\theta} = AV_0$, where $V_0$ is the zero-temperature value of
the ``phase stiffness'' (which sets the energy scale for the spatial variation 
of the superconducting phase) and $A$ is a number of  order 
unity.\cite{nature} $V_0$ may be expressed in terms of the superfluid density $n_s(T)$ 
or, equivalently, the experimentally-measured penetration depth $\lambda(T)$ 
at $T=0$:
\begin{equation}
V_0 = {\hbar^2 n_s(0)a \over 4m^*} 
=   {(\hbar c)^2 a \over 16\pi(e\lambda(0))^2}
\end{equation}
where $a$ is a length scale that depends on the dimensionality of the 
material. For a conventional superconductor such as Pb, $T_{\theta}$ is
about 10$^6$K, which implies that phase ordering occurs very close to the 
temperature at which pairing is established.\cite{nature} 

For the {\hts}, especially underdoped materials, $\Delta_0 /2 T_c > 1$, and it
varies with doping. The ratio $\Delta_0 /2T_c$ ranges from about 2 to 4 as a 
function of $x$. On the other hand, $T_{\theta}$ provides a quite 
good estimate of T$_c$ for the {\hts},\cite{nature} an estimate that can be 
improved by making a plausible generalization of the classical phase 
Hamiltonian.\cite{erica}  This behavior is qualitatively consistent with the
high-T$_c$ route to superconductivity in the 1DEG, as discussed above.

This phenomenology led us to 
conclude\cite{nature} that the spin gap observed in NMR and other 
experiments \cite{batlogg}  ({\it e.g.} as a
peak in  $(T_1T)^{-1}$ at a temperature T$_2^*$, where $T_1$ is the nuclear 
spin relaxation time) should be identified with a superconducting pseudogap
and {\it not} with a pseudogap associated with impending antiferromagnetic
order at zero doping. This identification is now supported by ARPES 
experiments on underdoped materials,\cite{arpesgap} that find a pseudogap 
above T$_c$ with the same shape and magnitude as the gap observed in the 
superconducting state. Also, in underdoped materials, the optical conductivity $\sigma_{ab}(\omega)$ 
in the $ab$-plane develops a pseudo-delta function, or a narrowing of the 
central ``Drude-like'' coherent peak above T$_c$.\cite{basov} 
Essentially all of the spectral weight moves downwards, 
which indicates the development of superconducting correlations. 

The existence of local superconducting correlations below $T^*_2$ indicates
that the amplitude of the order parameter is well established but there is no 
long-range phase coherence. This situation could, in principle, be realized
either by increasing $\Delta_0$ and elevating the pairing scale  
or by decreasing $n_s(0)$ and depressing the phase coherence scale as the
doping $x$ is decreased below its optimal value. 
Experimentally, as $x$ decreases, $\Delta_0$
varies very little (or even increases), whereas the superfluid density 
tends to zero as $x \rightarrow 0$. 
An increase in $\Delta_0$ would amount to 
a crossover to Bose-Einstein condensation, which also requires that
the chemical potential descend into the band or that the doped holes
form a separate band, both of which are
contradicted by ARPES experiments.\cite{arpes} 
In other words, the separation of the
temperature scales for pairing and phase coherence in underdoped {\hts} is
a consequence of the fact that the {\hts} are doped insulators; it is
{\it not} a crossover from BCS physics to Bose-Einstein condensation.

Another way of looking at the situation is to compare the superfluid density
$n_s(0)$ with the number of particles $n_P$ involved in pairing. In BCS theory,
at T=0, $n_P$ is of order $\Delta_0/E_F$ (where E$_F$ is the Fermi energy) 
and $n_s(0)$ is given by all the particles in the Fermi sea; {\it i.e.}
$n_p \ll n_s(0)$. For Bose condensation $n_P = n_s(0)$.
We shall argue that, in the
{\hts}, $n_P \gg n_s(0)$; most of the holes in the Fermi sea participate in 
the spin gap below T$_2^*$ but the superfluid density of the doped insulator 
is small. An intuitive although somewhat imprecise picture of the third 
possibility is provided by the hard-core dimer model \cite{dimer} in which
all the holes participate in dimers, but the mobile charge density is 
proportional to $x$.

\section{Spin gap proximity effect}

The existence of a charge-glass state \cite{rome1} in a substantial range
of doping in the {\hts} implies that
the dynamics of holes along the stripe is much faster than the fluctuation
dynamics of the stripe itself. Thus, on a finite length scale ($\sim 50 \AA$), 
an individual stripe may be regarded as a one-dimensional electron gas (1DEG) 
in an active environment of undoped spin regions between the stripes. Then it 
is appropriate to start out with a discussion of an extended 1DEG in which the 
singlet pair operator $P^{\dagger}$ may be written
\begin{equation}
P^{\dagger} = \psi^{\dagger}_{1 \ua} \psi^{\dagger}_{2 \da}  -
\psi^{\dagger}_{1 \da} \psi^{\dagger}_{2 \ua},  
\end{equation}
where $\psi^{\dagger}_{i, \sigma}$ creates a right-going ($i=1$) or
left-going ($i=2$) fermion with spin $\sigma$.
One route to superconductivity in the 1DEG is similar to the BCS many-body 
theory. At zero temperature in a gapless phase of the 1DEG, 
the correlation function $<P^{\dagger}(x,t) P(0,0)>$ is a 
power law with an exponent $K_c^{-1} + K_s$, where $K_c$ and $K_s$ are the 
critical exponent parameters for the charge and spin degrees of freedom and
specify the location of the system along lines of (quantum critical)
fixed points.\cite{review} For a non-interacting system, $K_c = K_s = 1$ so,
if $K_c^{-1} + K_s < 2$, pairing correlations are enhanced and pair hopping
between the different members of an array of 1DEG's will lead to a BCS-like
superconducting phase transition, in which pairing and phase coherence
develop at essentially the same temperature. Typically this is a 
low-temperature route to superconductivity and, like BCS theory, it requires 
an attractive interaction between the charge carriers 
({\it i.e.} $K_{c} > 1$, $K_{s} = 1$). 

However there is another route, that is much closer to the phenomenology
of the {\hts}. The fermion operators of a 1DEG may be expressed in terms
of Bose fields and their conjugate momenta ($\phi_c(x), \pi_c(x)$) and 
($\phi_s(x), \pi_s(x)$) corresponding to the charge and spin collective
modes respectively. In particular, the pair operator $P^{\dagger}$ becomes
\cite{review}
\begin{equation}
P^{\dagger} \sim e^{i\sqrt{2 \pi}\theta_c} \cos\big(\sqrt{2 \pi} \phi_s\big),
\end{equation}
where $\partial_x\theta_c \equiv \pi_c$. In other words, there is an 
{\it operator} relation in which the amplitude of the pairing operator depends 
on the spin fields only and the (superconducting) phase is a property of the 
charge degrees of freedom. Now, if the system acquires a spin gap, the 
amplitude $\cos\big(\sqrt{2 \pi} \phi_s\big)$ acquires a finite expectation
value, and superconductivity will appear when the charge degrees of freedom
become phase coherent. Below the spin-gap temperature, the critical exponent
of the pairing operator is given by $K_c^{-1}$, which can more easily fall 
below 2 and generate superconductivity for an array, because there is no 
contribution from $K_s$.\cite{review} More to the point the spin gap
temperature can be quite high, even in a single 1DEG, and it is generically
distinct from the phase ordering temperature.\cite{review,sak} Of course
phase order can only be established in a quasi-one dimensional system because,
in a simple 1DEG, it is destroyed by quantum fluctuations, even at zero 
temperature.

For an array of 1DEG's, a spin gap occurs only if there is 
an attractive interaction in the 
spin degrees of freedom. However, this is no longer true if the array is in
contact with an active (spin) environment, as in the stripe phases. We have
shown that pair hopping between the 1DEG and the environment will convey
a pre-existing spin gap from the environment to the 1DEG, or will generate
a spin gap in both the stripe and the environment, even for purely 
repulsive interactions.\cite{sgpe} A simple intuitive picture of this process
is as follows: The spin part of the singlet pair operator $P^{\dagger}$ 
on a stripe is
$\pm (\ua \da - \da \ua)/\sqrt 2$. On the other hand, locally, the spins in the 
environment have a N\'eel spin configuration $(\ua \da \ua \da \ua \da .....)$.
Then, by the exclusion principle, the amplitude for pair hopping between the
stripe and the environment has a (spin) factor $1/\sqrt 2$. However, pair 
hopping is enhanced by a factor $\sqrt 2$, and the kinetic energy lowered 
if the spins in the environment also form singlets.
Note that the sign of the singlet wave functions in the environment must
be chosen to maximize the overall hopping amplitude of the pairs, as the
phase $\theta_c$ varies along a stripe. This corresponds to the composite 
order parameter that appears in the quantum field theory treatment of the 
problem.\cite{sgpe} In principle, this process may not lead to a gap for
all of the spins in the environment in the normal state. However, once
pair hopping between the stripes becomes coherent, the remaining spins
will acquire a gap via the spin gap proximity effect.\cite{sgpe}

This mechanism of {\hty} also avoids problem of the strong Coulomb 
interaction because it involves pairing of neutral femions, or spinons,
that are known to exist in the one-dimensional electron gas.\cite{review}  
It allows a spin gap with a range of one lattice spacing in the environment
and about two lattice spacings on a stripe.

Not only does this route to superconductivity correspond closely to the
phenomenology of the {\hts} but it also works for a short stripe. It is 
well known, {\it e.g.} from an analysis of numerical calculations, that,
if the length scale associated with the spin gap is short compared to the
length of a stripe, then the calculation for an infinite system is a good
approximation for the finite system. Furthermore, once the spin degrees of
freedom are frozen in this way, the remaining Hamiltonian corresponds to
a phase-number model that we have used to analyse the effects of quantum
phase fluctuations.\cite{badmetals} Superconductivity appears when the different
stripes become phase coherent, and {\it the superconducting coherence length 
is given by the spacing between stripes and not by the range of the pair wave
function as in BCS theory.} A consequence is that, in the superconducting 
state, the radius of a vortex core should have a very weak temperature 
dependence, and that the core should be an essentially undoped region with a
spin gap. Both of these conclusions are supported by 
experiment.\cite{ubc,geneva}

\section{Momentum space}

So far we have dicussed the consequences of stripes in real space. But
ARPES experiments show that the {\hts} have a ``Fermi surface'' 
even though there are no well-defined quasiparticles. Therefore it is 
appropriate to ask how this physics is realized in momentum space. We have 
calculated the spectral function of a simplified stripe model and have found 
a reasonable correspondence with the ARPES experiments.\cite{markku} 
The spin and charge wave vectors transverse to vertical stripes span the
``Fermi surface'' in the neighborhood of the points $(\pm \pi,0)$ 
and give rise to {\it regions} of degenerate states. Horizontal stripes have 
the same effect in the neighborhood of $(0, \pm \pi)$.  These are 
indeed the regions in which {\hty} originates.\cite{arpes3} In practice, 
these regions are connected by arcs that are approximately 45$^{\circ}$  
sections of a circle. Along these arcs, stripe wave vectors span the 
``Fermi surface'' at isolated points at most. Therefore the arc must become 
aware of the stripes by many-body effects such as the scattering of a pair of 
particles with total momentum zero into the regions near the $\bar M$ points
$(\pm \pi,0)$ and $(0, \pm \pi)$. This implies that the spin gap should
spread over the arcs as the system is cooled below the spin-gap temperature,
which is consistent with ARPES observations.\cite{arpes3}

\subsection{Symmetry of the order parameter}

The momentum space picture also has consequences for the symmetry of the
order parameter. The regions near to $(\pm \pi,0)$ and $(0, \pm \pi)$
communicate with each other via the arcs of the ``Fermi surface'', and the 
relative phase of these regions must be chosen to maximize the amplitude of 
the order parameter along the arcs. As mentioned above, experimentally, the 
range of the gap 
function is nearest neighbor in real space for optimal doping, corresponding 
to the $d$-wave $\cos k_x - \cos k_y$ or the extended $s$-wave 
$\cos k_x + \cos k_y$. Evidently the amplitude of the extended $s$-wave 
vanishes at the $\bar M$ points, so the $d$-wave order parameter has the 
greater condensation energy.\cite{vjeprl}

This view of the origin of the symmetry of the order parameter leads to a
number of interesting consequences. First of all, the existence of a
nearest-neighbor gap function along the arcs of the ``Fermi surface'' 
suggests that the arcs correspond to the regions between stripes. Secondly,
for the second and third neighbor components of the gap function, the 
amplitudes of the $d$-wave components ($\sin k_x sin k_y$ and 
$\cos 2k_x - \cos 2ky$) vanish at the $\bar M$ points but the amplitudes of 
the $s$-wave components ($\cos k_x cos k_y$ and $\cos 2k_x + \cos 2ky$) are 
maximized. In certain circumstances, these $s$-wave components of the order
parameter could either mix with the $d$-wave component (breaking time-reversal 
symmetry or lattice-rotational symmetry) or even become dominant. 
There is evidence from tunnelling 
spectroscopy that order parameter mixing is induced in surfaces of 
{\YBCO}.\cite{lhg} An $s$-wave order parameter or component of the order 
parameter might also appear in overdoped materials, where the
stripe structure is breaking up: the increased meandering of the stripes
will tend to mix the short-range gap function of the environment with the
longer-range gap function on the stripes.

{\bf Acknowledgements:}  We would like to aknowledge frequent discussions of
the physics of {\hts} with J.~Tranquada. 
This work was supported at UCLA by the National Science Foundation grant 
number DMR93-12606 and, at Brookhaven,  by the Division of Materials Sciences,
U. S. Department of Energy under contract No. DE-AC02-98CH10886.

\end{document}